\documentclass{article}

\usepackage{amsmath,amssymb,booktabs,array}
\usepackage[normalem]{ulem}
\usepackage{bm}
\usepackage{xcolor}
\usepackage{graphicx}
\usepackage{microtype}
\usepackage{url}
\usepackage{hyperref}
\usepackage[round]{natbib}
\usepackage{fancyvrb}
\usepackage{arxiv}

\newcommand{\proglang}[1]{\textsf{#1}}
\newcommand{\pkg}[1]{\textbf{#1}}
\newcommand{\code}[1]{\texttt{#1}}
\newcommand{\fct}[1]{\texttt{#1()}}

\newcommand{\email}[1]{\href{mailto:#1}{\texttt{#1}}}

\newenvironment{CodeChunk}{\par\vspace{4pt}}{\par\vspace{4pt}}

\DefineVerbatimEnvironment{CodeInput}{Verbatim}{%
  fontsize=\small, baselinestretch=1, xleftmargin=4mm,
  frame=leftline, framerule=0.5pt, framesep=4pt}

\DefineVerbatimEnvironment{CodeOutput}{Verbatim}{%
  fontsize=\small, baselinestretch=1, xleftmargin=4mm}

\DefineVerbatimEnvironment{Code}{Verbatim}{%
  fontsize=\small, baselinestretch=1, xleftmargin=4mm,
  frame=leftline, framerule=0.5pt, framesep=4pt}

\title{Statistical Inference for Gaussian Kernel Robust Regression
       with the \pkg{gkrreg} Package}

\author{Marcelo Rodrigo Portela Ferreira \\
  Departamento de Estat\'istica \\
  Universidade Federal da Para\'iba \\
  Jo\~ao Pessoa, PB 58051-900, Brazil \\
  \email{marcelo@de.ufpb.br}
  \And
  Eufr\'asio de Andrade Lima Neto \\
  Departamento de Estat\'istica \\
  Universidade Federal da Para\'iba \\
  Jo\~ao Pessoa, PB 58051-900, Brazil \\
  \email{eufrasio@de.ufpb.br}
}

\renewcommand{\shorttitle}{Inference for GKRReg in \proglang{R}}

\hypersetup{
  pdftitle={Statistical Inference for Gaussian Kernel Robust Regression
            with the gkrreg Package},
  pdfsubject={stat.ME, stat.CO},
  pdfauthor={Marcelo R. P. Ferreira, Eufr\'asio de A. Lima Neto},
  pdfkeywords={robust regression, Gaussian kernel, sandwich estimator,
               bootstrap, redescending M-estimator, R},
}

\begin{document}
\maketitle

\begin{abstract}
The Gaussian Kernel Robust Regression method (GKRReg) is a robust regression estimator that iteratively re-weights observations via a Gaussian kernel so that outliers and leverage points receive near-zero weight, with convergence of the estimation algorithm theoretically guaranteed. Despite a thorough study of estimation, the original work leaves open the problem of statistical inference for the regression coefficients. We fill this gap with three contributions. First, we formally establish that GKRReg belongs to the family of \emph{redescending} $M$-estimators, providing the theoretical foundation for the inferential procedures that follow. Second, we derive a \emph{closed-form analytic sandwich variance estimator} based on the theory of generalised $M$-estimators, corresponding to the HC0 class of heteroskedasticity-robust covariance matrices; we show that a finite-sample correction analogous to HC3 requires the weighted hat matrix of the converged IRWLS step, and identify this as a direction for future work. Third, we propose a \emph{pairs bootstrap that re-estimates the kernel width hyper-parameter} $\hat\gamma^2$ on every replicate, capturing variability that the sandwich ignores. All procedures are implemented in the \proglang{R} package \pkg{gkrreg}, which also provides four estimators for $\gamma^2$ and an automatic data-driven selection procedure, comprehensive diagnostic plots, and six real datasets from the robust regression literature. Applications to real data sets and comparison with traditional robust regression models highlight the potential of the GKRReg and the usability of the \proglang{R} package.
\end{abstract}



\section[Introduction]{Introduction} \label{sec:intro}

Outliers and leverage points are a pervasive challenge in applied regression analysis.  Classical ordinary least squares (OLS) is highly sensitive to such observations because the squared-error loss assigns unbounded influence to residuals of any size.  A rich literature has developed robust alternatives, including $M$-estimators \citep{Huber:1964}, $S$-estimators \citep{Rousseeuw+Yohai:1984} and $MM$-estimators \citep{Yohai:1987}. Implementations for \proglang{R} \citep{R} include the \pkg{MASS} package \citep{Venables+Ripley:2002} and the \pkg{robustbase} package \citep{Maechler+:2023}. Bootstrap inference for robust estimators such as $S$- and $MM$-estimators is available in the \pkg{FRB} package \citep{VanAelst+Willems:2013}, which implements the fast and robust bootstrap for multivariate regression, principal component analysis and Hotelling tests.

\citet{DeCarvalho+LimaNeto+Ferreira:2017} proposed the Exponential-Type
Kernel function based Robust Regression (ETKRR), which re-weights observations using exponential-type kernel functions within an iteratively re-weighted least squares (IRWLS) framework.  The Gaussian specialisation, which we refer to as Gaussian Kernel Robust Regression (GKRReg), offers a particularly clean mathematical structure since the Gaussian kernel $G(a,b)=\exp(-(a-b)^2/\gamma^2)$ is differentiable everywhere and its gradient with respect to the fitted value
has the closed form $\partial G / \partial\mu_i = (2e_i/\gamma^2)\,G$, enabling analytic derivations that are not available for kernels without this property. Unlike the Tukey bisquare estimator, which applies a hard cutoff beyond which observations receive exactly zero weight, GKRReg
implements a soft, exponentially decaying down-weighting governed by the kernel width $\gamma^2$, i.e., observations with larger residuals contribute negligibly to $\hat{\boldsymbol\beta}$, but the transition from full influence to near-zero influence is smooth and differentiable, a property that is essential for the analytic sandwich derivations presented in Section~\ref{sec:inference}. Convergence of the IRWLS algorithm is theoretically guaranteed (Propositions 4.1--4.2 of \citealt{DeCarvalho+LimaNeto+Ferreira:2017}).

Despite a thorough experimental evaluation comparing GKRReg against OLS and classical robust estimators (M, MM and S) under several simulated contamination scenarios as well as on real datasets~\citep{DeCarvalho+LimaNeto+Ferreira:2017}, the original paper leaves open the question of statistical inference. Specifically, no sampling distribution, standard errors, confidence intervals or hypothesis tests for $\hat{\boldsymbol\beta}$ were provided. This gap is practically significant. The weighted least squares covariance matrix $(\mathbf{X}^\top\hat{\mathbf{K}}\mathbf{X})^{-1}$, obtained at the final
IRWLS step, treats the kernel weight matrix  $\hat{\mathbf{K}}$ as fixed and therefore systematically underestimates the true variance of $\hat{\boldsymbol\beta}$.

Statistical inference for robust estimators is non-trivial in general because the estimating equations are non-linear in the parameters and the influence function is bounded, so standard maximum likelihood theory does not apply directly. For $M$- and $MM$-estimators, the sandwich variance estimator based on the theory of generalised $M$-estimators~\citep{Huber:1981,Stefanski+Boos:2002} provides a tractable analytic solution, and fast bootstrap approximations are available~\citep{VanAelst+Willems:2013}. For GKRReg, two additional difficulties arise. First, the kernel weights $k_{ii} = G(y_i, \mathbf{x}_i^\top\boldsymbol\beta)$ depend implicitly on $\boldsymbol\beta$, so the bread matrix of the sandwich estimator requires differentiating the weights with respect to the parameters --- a step that is non-trivial but, as we show, has a closed form for the Gaussian kernel. Second, the kernel width hyper-parameter $\gamma^2$ is estimated from the data in a preliminary step and treated as fixed throughout the IRWLS algorithm; the uncertainty in $\hat\gamma^2$ is therefore not captured by the sandwich estimator and must be accounted for by a bootstrap that re-estimates $\gamma^2$ on every replicate.

The present paper makes three theoretical contributions. First, we formally establish that GKRReg is a \emph{redescending} $M$-estimator in the sense of \citet{Maronna+Martin+Yohai:2006}, providing the theoretical foundation for the inference procedures that follow. Second, we derive closed-form expressions for the analytic sandwich variance estimator and show that the result belongs to the HC0 class of heteroskedasticity-robust covariance matrices \citep{White:1982}; we discuss finite-sample corrections analogous to HC2/HC3~\citep{MacKinnon+White:1985,Davidson+MacKinnon:1993} and why they require additional development for the GKRReg context. Third, we propose a pairs bootstrap~\citep{Efron:1979,Efron+Tibshirani:1993} that re-estimates $\hat\gamma^2$ on every replicate, capturing variability that the sandwich ignores. Additionally, but no less important, all procedures are implemented in the \proglang{R} package \pkg{gkrreg} \citep{gkrreg:2025}, available from the Comprehensive \proglang{R} Archive Network (CRAN) at
\url{https://CRAN.R-project.org/package=gkrreg}.

The remainder of the paper is organised as follows. Section~\ref{sec:method} reviews the GKRReg method, establishes its connection to redescending $M$-estimators, and describes the four estimators for $\gamma^2$. Section~\ref{sec:inference} derives the sandwich variance estimator, the bootstrap procedure and discusses the relationship to the HC family. Section~\ref{sec:package} presents the \pkg{gkrreg} package. Section~\ref{sec:illustrations} illustrates the methods on six real datasets. Section~\ref{sec:summary} concludes the paper highlighting the most important findings.


\section[The GKRReg method]{The GKRReg Method} \label{sec:method}

\subsection[Model and objective function]{Model and Objective Function}
\label{subsec:model}

Consider the linear model
$y_i = \mathbf{x}_i^\top\boldsymbol\beta + \varepsilon_i$,
$i = 1,\ldots,n$, where $y_i$ represents the response variable, $\mathbf{x}_i$ is a $p \times 1$ vector within the independent variables, $\boldsymbol\beta$ is a $p \times 1$ vector of unknown parameters, and $\varepsilon_i$ are i.i.d. errors with zero mean.
GKRReg minimises
\begin{equation}
  S(\boldsymbol\beta)
  = 2\sum_{i=1}^n\bigl[1 - G(y_i,\hat\mu_i)\bigr],
  \qquad
  G(a,b) = \exp\!\Bigl(-\tfrac{(a-b)^2}{\gamma^2}\Bigr),
  \label{eq:objective}
\end{equation}
where $\hat\mu_i = \mathbf{x}_i^\top\boldsymbol\beta$ is the conditional expectation of $Y|X=x$ and $\gamma^2>0$ is
the kernel width hyper-parameter. Let $r_i = y_i - \hat{\mu}_i$ be the ordinary residual, the corresponding $\rho$-function is given by
\begin{equation}
  \rho_{\mathrm{GK}}(r_i) = 1 - \exp\Bigl(-\tfrac{r^2_i}{\gamma^2}\Bigr).
  \label{eq:rho}
\end{equation}
Setting $\partial S/\partial\boldsymbol\beta = \mathbf{0}$ and using the
identity
\begin{equation}
  \frac{\partial G(y_i,\mu_i)}{\partial\mu_i}
  = \frac{2(y_i-\mu_i)}{\gamma^2}\,G(y_i,\mu_i),
  \label{eq:kernel_deriv}
\end{equation}
yields IRWLS normal equations with kernel weights
$k_{ii} = G(y_i,\hat\mu_i)\in(0,1]$, i.e.,
\begin{equation}
  \frac{\partial S}{\partial \boldsymbol{\beta}}
  = \frac{2}{\gamma^2}
    \sum_{i=1}^{n} G(y_i, \hat{\mu}_i)\,(y_i - \hat{\mu}_i)\,\mathbf{x}_i
  = \mathbf{0}.
  \label{eq:normal_equations}
\end{equation}

The IRWLS algorithm that minimises~\eqref{eq:objective} proceeds as follows. Given a preliminary estimate $\hat\gamma^2$ obtained by one of the estimators described in Section~\ref{subsec:gamma}, the iterations are:
\begin{enumerate}
  \item \textbf{Initialisation.} Compute the OLS estimate
    $\hat{\boldsymbol\beta}^{(0)} =
    (\mathbf{X}^\top\mathbf{X})^{-1}\mathbf{X}^\top\mathbf{y}$
    and set $t = 1$.
  \item \textbf{Weights.} Compute residuals
    $r_i^{(t-1)} = y_i - \mathbf{x}_i^\top\hat{\boldsymbol\beta}^{(t-1)}$
    and kernel weights
    $$k_{ii}^{\textcolor{blue}{(t)}} = G(y_i, \mathbf{x}_i^\top\hat{\boldsymbol\beta}^{(t-1)})
    = \exp\!\bigl(-(r_i^{(t-1)})^2/\hat\gamma^2\bigr) \in (0, 1]$$.
  \item \textbf{Update.} Solve the weighted least squares problem
    \begin{equation}
      \hat{\boldsymbol\beta}^{(t)}
      = \bigl(\mathbf{X}^\top\hat{\mathbf{K}}^{(t)}\mathbf{X}\bigr)^{-1}
        \mathbf{X}^\top\hat{\mathbf{K}}^{(t)}\mathbf{y},
      \label{eq:irwls_update}
    \end{equation}
    where $\hat{\mathbf{K}}^{(t)} =
    \operatorname{diag}(k_{11}^{(t)},\ldots,k_{nn}^{(t)})$.
  \item \textbf{Convergence check.} If
    $\|\hat{\boldsymbol\beta}^{(t)} - \hat{\boldsymbol\beta}^{(t-1)}\|_\infty
    < \varepsilon$ (default $\varepsilon = 10^{-10}$ in \pkg{gkrreg}),
    set $\hat{\boldsymbol\beta} = \hat{\boldsymbol\beta}^{(t)}$ and stop;
    otherwise set $t \leftarrow t+1$ and return to Step~2.
\end{enumerate}

At the initialisation step, $\mathbf{K}^{(0)} = I_n$, meaning that all observations (including outliers) receive the same weight (importance) for the parameter estimates $\hat{\boldsymbol\beta}^{(0)}$, which therefore coincide with the usual OLS estimator. However, at the model weight step, $\mathbf{K}$ is updated using the ordinary residuals, which helps to discriminate well-fitted observations from bad ones in terms of their importance for the estimates of $\boldsymbol\beta$. \citet{DeCarvalho+LimaNeto+Ferreira:2017} prove that
$S(\hat{\boldsymbol\beta}^{(t)}) \le S(\hat{\boldsymbol\beta}^{(t-1)})$ at every iteration and that the sequence $\{\hat{\boldsymbol\beta}^{(t)}\}$ converges to a fixed point of the IRWLS map (Propositions~4.1--4.2 therein), guaranteeing that the algorithm terminates in a finite number of steps for any $\varepsilon > 0$. In summary, the algorithm starts from OLS and alternates between WLS updates and weight recomputations until convergence. At convergence, Equations~\eqref{eq:normal_equations} can be written compactly as the estimating equations of a generalised $M$-estimator
\citep{Huber:1981,Stefanski+Boos:2002}:
\begin{equation}
  \boldsymbol{\Psi}(\boldsymbol{\hat{\beta}})
  \;=\;
  \sum_{i=1}^{n} \psi(\mathbf{x}_i, y_i;\, \boldsymbol{\hat{\beta}})
  \;=\;
  \sum_{i=1}^{n} k_{ii}(\boldsymbol{\hat{\beta}})\,
  \mathbf{x}_i\,(y_i - \mathbf{x}_i^{\top}\boldsymbol{\hat{\beta}})
  \;=\; \mathbf{0},
  \label{eq:estimating_equations}
\end{equation}
where $k_{ii}(\boldsymbol{\hat{\beta}}) = G(y_i,\mathbf{x}_i^{\top}\boldsymbol{\hat{\beta}})$ makes the weights implicitly dependent on $\boldsymbol{\hat{\beta}}$.  This dependence is critical, from an inference point of view, since the weighted least
squares covariance matrix $(\mathbf{X}^{\top}\hat{\mathbf{K}}\mathbf{X})^{-1}$ obtained at the final IRWLS step treats $\hat{\mathbf{K}}$ as fixed, ignoring the variability introduced by estimating both the weights and the
hyper-parameter $\gamma^2$. Consequently, the true variance of $\hat{\boldsymbol{\beta}}$ is systematically underestimated
and the inference based on it is invalid.

\subsection[GKRReg as a redescending M-estimator]{GKRReg as a Redescending $M$-Estimator} \label{subsec:redescending}

Following Definition~2.1 of \citet{Maronna+Martin+Yohai:2006}, a
$\rho$-function defines a valid $M$-estimator if it satisfies:
\begin{description}
    \item[(P1)]~$\rho(x)$ is nondecreasing in $|x|$;
    \item[(P2)]~$\rho(0)=0$;
    \item[(P3)]~$\rho(x)$ is increasing for $x>0$ with $\rho(x)<\rho(\infty)$;\item[(P4)]~$\rho$ is bounded, $\rho(\infty)<\infty$.
\end{description}

Let $r \in \mathbb{R}$ be the ordinary residual, we verify each property for $\rho_{\mathrm{GK}}$, given by Eq.~\eqref{eq:rho}, as follows:
\begin{description}
\item[(P1)] As $|r|$ increases, $r^2$ increases. Since $\gamma^2 > 0$, the exponent term $-\frac{r^2}{\gamma^2}$ decreases. Consequently, $\exp\left(-\frac{r^2}{\gamma^2}\right)$ decreases, which means $1 - \exp\left(-\frac{r^2}{\gamma^2} \right)$ increases. Therefore, $\rho_{\mathrm{GK}}(r)$ is nondecreasing in $|r|$;
\item[(P2)] Substituting $r = 0$ into the expression:
    $$\rho_{\mathrm{GK}}(0) = 1 - \exp\left( -\frac{0^2}{\gamma^2} \right) = 1 - \exp(0) = 1 - 1 = 0.$$
    Thus, this property is satisfied;
\item[(P3)] $\rho'_{\mathrm{GK}}(r) = d\rho_{\mathrm{GK}}/{dr} = (2r/\gamma^2) \exp(-{r^2}/{\gamma^2})$. For $r > 0$ and $\gamma^2 > 0$, we have $\rho'_{\mathrm{GK}}(r) > 0$, confirming that the function is strictly increasing for $r > 0$. Furthermore, as $|r| \to \infty$, $\exp(-{r^2}/{\gamma^2}) \to 0$, so $\rho_{\mathrm{GK}}(\infty) = 1$. Since $\exp(-{r^2}/{\gamma^2}) > 0$ for all finite $r$, it follows that $\rho_{\mathrm{GK}}(r) < 1$ for all finite $r$. Thus, $\rho_{\mathrm{GK}}(r) < \rho_{\mathrm{GK}}(\infty)$
\item[(P4)] $\lim_{|r|\to\infty}\rho_{\mathrm{GK}}(r)=1<\infty$. As shown above, the limit of the function as $r \to \infty$ is:
$$\lim_{r \to \infty} \left( 1 - \exp\left( -\frac{r^2}{\gamma^2} \right) \right) = 1 - 0 = 1.$$
The function is bounded and correctly normalized to 1.
\end{description}

The influence function is $\psi_{\mathrm{GK}}(r) \propto
r\,\exp(-r^2/\gamma^2)$.
Because $\exp(-r^2/\gamma^2)\to 0$ faster than $r$ grows,
$\lim_{|r|\to\infty}\psi_{\mathrm{GK}}(r) = 0$: the influence function
\emph{redescends} to zero.
GKRReg therefore belongs to the class of \emph{redescending}
$M$-estimators.
Geometrically, $\psi_{\mathrm{GK}}$ attains its maximum at
$|r|=\gamma/\sqrt{2}$ and then decays exponentially, so observations
with residuals much larger than $\gamma/\sqrt{2}$ are effectively ignored.

A redescending $M$-estimator is one whose influence function
$\psi(r) = \rho'(r)$ eventually returns to zero as $|r| \to \infty$,
in contrast to monotone $M$-estimators such as the Huber estimator, whose
influence function is bounded but remains strictly positive for all large
residuals. The practical consequence of this difference is that, while a
Huber-type estimator assigns reduced but non-zero influence to outlying
observations, a redescending estimator assigns \emph{zero} asymptotic
influence to observations whose residuals exceed a certain threshold,
effectively excluding them from the parameter estimation.
For GKRReg, this threshold is not a hard cutoff as in the Tukey bisquare
estimator, but a soft one guided by the kernel width-hyperparameter $\gamma^2$. Thus, the
influence function $\psi_{\mathrm{GK}}(r) \propto r\exp(-r^2/\gamma^2)$
attains its maximum at $|r| = \gamma/\sqrt{2}$ and decays exponentially
thereafter, so that observations with residuals much larger than
$\gamma/\sqrt{2}$ contribute negligibly to $\hat{\boldsymbol\beta}$.
In terms of the IRWLS algorithm, this translates directly into kernel
weights $k_{ii} = \exp(-r_i^2/\hat\gamma^2) \approx 0$ for outlying
observations, which are therefore down-weighted to near zero at convergence without requiring the analyst to identify or remove them manually.

\subsection[Comparison with the bisquare estimator]{Comparison with the Bisquare Estimator} \label{subsec:bisquare}

Both GKRReg and the Tukey bisquare estimator are redescending but differ fundamentally in their rejection mechanism. Table~\ref{tab:bisquare} summarises the key differences.
\begin{table}[h!]
\centering
\caption{\label{tab:bisquare}Comparison between the bisquare $M$-estimator and GKRReg.}
\begin{tabular}{lp{5cm}p{5cm}}
\toprule
Feature & Bisquare & GKRReg \\
\midrule
$\psi$-function &
  $r[1-(r/c)^2]^2\mathbf{1}(|r|\le c)$ &
  $(r/\gamma^2)\exp(-r^2/\gamma^2)$ \\[4pt]
Support &
  Compact: $\psi\equiv 0$ for $|r|>c$ &
  Infinite: $\psi\to 0$ but $\psi>0$ for all finite $r$ \\[4pt]
Outlier rejection &
  Hard: zero weight beyond $c$ &
  Soft: exponentially small but positive weight \\[4pt]
Convergence &
  Requires robust initialisation &
  Guaranteed \citep{DeCarvalho+LimaNeto+Ferreira:2017} \\[4pt]
Tuning &
  $c$ (fixed cutoff) &
  $\gamma^2$ (estimated from data) \\
\bottomrule
\end{tabular}
\end{table}
The soft rejection of GKRReg has a direct consequence for inference. Since $k_{ii}>0$ for all observations and depends implicitly on $\boldsymbol\beta$, the sandwich derivation requires the full gradient $\partial k_{ii}/\partial\boldsymbol\beta$ (Section~\ref{subsec:sandwich_AB}),
which has no analogue in the bisquare case.

\subsection[Width hyper-parameter estimators]{Width Hyper-Parameter Estimators} \label{subsec:gamma}

The performance of GKRReg depends on $\gamma^2$.
Four estimators are available in \pkg{gkrreg}, all evaluated on OLS
residuals $r_i^{\mathrm{OLS}}$ before IRWLS begins:
\begin{description}
  \item[S1~(Caputo's)] Mean of the 0.1 and 0.9 quantiles of
    $(y_i-\hat\mu_j^{\mathrm{OLS}})^2$, $\forall i \neq j$ on a random sub-sample of size $n^*=\lfloor\alpha n\rfloor$ (default $\alpha=0.5$)\citep{Caputo+:2002}. Recommended for clean data.
  \item[S2 (Pairwise median)] Median of $(y_i-\hat\mu_j^{\mathrm{OLS}})^2$,
    $\forall i\ne j$.  Recommended for Y-space outliers $\le 10\%$ and X-space
    outliers $\le 15\%$.
  \item[S3 (Residual variance)]
    $\hat\gamma^2 = \sum(y_i-\hat\mu_j^{\mathrm{OLS}})^2/(n-p-1)$.
    Recommended for Y-space outliers $\ge 15\%$ and leverage points.
  \item[S4 (AICc bandwidth)] $\hat\gamma^2 = h_{\mathrm{AICc}}^2$, where
    $h_{\mathrm{AICc}}$ is the bandwidth minimising the corrected Akaike
    information criterion in a nonparametric regression of $y$ on
    $\hat\mu^{\mathrm{OLS}}$, via \fct{sm::h.select} from the \pkg{sm}
    package \citep{Bowman+Azzalini:1997}.
    Recommended for large samples ($n\ge 200$).
\end{description}
When the appropriate estimator is not obvious, \code{sigma\_method = "auto"} selects among S1--S4 by out-of-bag bootstrap mean squared prediction error (Section~\ref{subsec:auto}).

A natural question arising from the GKRReg formulation is whether the
kernel width hyper-parameter $\gamma^2$ could be estimated jointly with
$\boldsymbol\beta$ within the IRWLS algorithm itself, eliminating the need
for a preliminary estimator. A careful analysis of the objective function $S(\boldsymbol\beta, \gamma^2)$ shows that this is not possible in a straightforward way: $S$ is strictly monotone decreasing in $\gamma^2$ for any fixed $\boldsymbol\beta$ and any dataset with non-zero residuals, so the unconstrained minimiser with respect to $\gamma^2$ is the degenerate solution $\gamma^2 \to 0^+$, which collapses the Gaussian kernel to a point mass and assigns zero weight to every observation with a non-zero residual. This reinforces the role of $\gamma^2$ as a regularisation hyper-parameter rather than a model parameter. Minimising the loss with respect to it is ill-posed without an additional constraint, just as minimising a penalised regression loss with respect to the penalty parameter $\lambda$ is ill-posed without a model selection criterion. The estimators S1--S4 described in this section can therefore be understood not as approximations to an infeasible analytical solution, but as principled external calibration strategies that anchor $\gamma^2$ to the scale of the residuals before the IRWLS iterations begin.


\section[Statistical inference]{Statistical Inference} \label{sec:inference}

\subsection[Estimation equations at convergence]{Estimation Equations at Convergence} \label{subsec:estim_eq}


As established at the end of Section~\ref{subsec:model}, the converged GKRReg
normal equations can be written as a generalised $M$-estimator
\citep{Huber:1981,Stefanski+Boos:2002}:
\begin{equation}
  \boldsymbol\Psi(\boldsymbol\beta)
  = \sum_{i=1}^n k_{ii}(\boldsymbol\beta)\,
    \mathbf{x}_i\,(y_i-\mathbf{x}_i^\top\boldsymbol\beta)
  = \mathbf{0},
  \label{eq:estim_eq}
\end{equation}
where $k_{ii}(\boldsymbol\beta) = G(y_i,\mathbf{x}_i^\top\boldsymbol\beta)$.
The implicit dependence of $k_{ii}$ on $\boldsymbol\beta$, and through it
on $\hat\gamma^2$, means that neither the weights nor the hyper-parameter
can be treated as fixed when assessing the variability of
$\hat{\boldsymbol\beta}$.
We now derive the asymptotic distribution of $\hat{\boldsymbol\beta}$ that
properly accounts for this structure.

\subsection[Sandwich variance estimator]{Sandwich Variance Estimator}
\label{subsec:sandwich}

\subsubsection[Asymptotic theory]{Asymptotic Theory}

Under standard regularity conditions for $M$-estimators
\citep{Huber:1967,White:1982,Newey+McFadden:1994}, i.e., measurability, twice
continuous differentiability in $\boldsymbol\beta$, and
$\mathbf{E}[\|\psi_i\|^2]<\infty$, the solution of~\eqref{eq:estim_eq}
satisfies
\begin{equation}
  \sqrt{n}\,(\hat{\boldsymbol\beta}-\boldsymbol\beta_0)
  \;\xrightarrow{d}\;
  \mathcal{N}\!\bigl(\mathbf{0},\;
    \mathbf{A}^{-1}\mathbf{B}(\mathbf{A}^{-1})^\top
  \bigr),
  \label{eq:sandwich_asymp}
\end{equation}
with bread $\mathbf{A} = -\mathbf{E}[\partial\psi_i/\partial\boldsymbol\beta^\top]$
and meat $\mathbf{B} = \mathbf{E}[\psi_i\psi_i^\top]$
\citep{White:1982,Liang+Zeger:1986}.

\subsubsection[Closed-form expressions]{Closed-Form Expressions for
               $\hat{\mathbf{A}}$ and $\hat{\mathbf{B}}$}
\label{subsec:sandwich_AB}

Differentiating $\psi_i = k_{ii}(\boldsymbol\beta)\,\mathbf{x}_i e_i$ and
applying~\eqref{eq:kernel_deriv}:
\begin{equation}
  \frac{\partial k_{ii}}{\partial\boldsymbol\beta}
  = \frac{2(y_i - \mu_i)}{\gamma^2}\,k_{ii}\,\mathbf{x}_i.
  \label{eq:dk_dbeta}
\end{equation}
The product rule then gives
\begin{equation}
  \frac{\partial\psi_i}{\partial\boldsymbol\beta^\top}
  = k_{ii}\,\mathbf{x}_i\mathbf{x}_i^\top
    \!\left(\frac{2(y_i - \mu_i)^2}{\gamma^2}-1\right).
  \label{eq:grad_psi}
\end{equation}
The sample estimators evaluated at $\hat{\boldsymbol\beta}$ are
\begin{align}
  \hat{\mathbf{A}}
  &= \frac{1}{n}\,\mathbf{X}^\top
    \operatorname{diag}\!\left[k_{ii}\!\left(1-\frac{2 r_i^2}{\hat\gamma^2}
    \right)\right]\mathbf{X},
  \label{eq:Ahat}\\[4pt]
  \hat{\mathbf{B}}
  &= \frac{1}{n}\,\mathbf{X}^\top
    \operatorname{diag}\!\left(k_{ii}^2 r_i^2\right)\mathbf{X},
  \label{eq:Bhat}
\end{align}
where $ r_i = y_i - \mathbf{x}_i^\top\hat{\boldsymbol\beta}$.
The sandwich covariance estimator is
\begin{equation}
  \widehat{\mathrm{Var}}(\hat{\boldsymbol\beta})
  = \frac{1}{n}\,\hat{\mathbf{A}}^{-1}\hat{\mathbf{B}}\hat{\mathbf{A}}^{-1}.
  \label{eq:sandwich}
\end{equation}
Standard errors, Wald $z$-statistics $z_j=\hat\beta_j/\widehat{\mathrm{se}}_j$,
two-sided $p$-values $p_j = 2\Phi(-|z_j|)$, and $(1-\alpha)$ confidence
intervals $\hat\beta_j \pm z_{\alpha/2}\widehat{\mathrm{se}}_j$ follow
directly.

\subsubsection[Relationship to the HC family]{Relationship to the HC Family}
\label{subsec:hc}

Estimator~\eqref{eq:sandwich} uses the squared residuals $r_i^2$ in the
meat~\eqref{eq:Bhat} without finite-sample correction, corresponding to the
HC0 estimator of \citet{White:1982}.
The HC1--HC4 corrections \citep{MacKinnon+White:1985,Davidson+MacKinnon:1993,
Cribari-Neto:2004} inflate residuals by factors involving the leverage
$h_{ii} = [\mathbf{X}(\mathbf{X}^\top\mathbf{X})^{-1}\mathbf{X}^\top]_{ii}$.
For GKRReg, the analogue of the leverage is the diagonal of the weighted
hat matrix
\begin{equation}
  \mathbf{H}_{\mathrm{GK}}
  = \mathbf{X}(\mathbf{X}^\top\hat{\mathbf{K}}\mathbf{X})^{-1}
    \mathbf{X}^\top\hat{\mathbf{K}},
  \label{eq:hat_matrix}
\end{equation}
which depends on $\hat{\mathbf{K}}$ and hence on $\hat{\boldsymbol\beta}$
and $\hat\gamma^2$.
A GKRReg-HC3 correction using $k_{ii}^2r_i^2/(1-h_{\mathrm{GK},ii})^2$
is theoretically well-defined but requires forming $\mathbf{H}_{\mathrm{GK}}$
at each IRWLS iterate.
We implement HC0 in \pkg{gkrreg} and identify GKRReg-HC3 as a direction for
future work.

\subsubsection[Limitations]{Limitations}

The sandwich estimator has two further limitations to GKRReg.
First, as a first-order asymptotic result, it may be unreliable for
small~$n$.
Second, $\hat\gamma^2$ is treated as fixed in~(\ref{eq:Ahat})--(\ref{eq:Bhat}),
so its sampling variability is not propagated into the standard errors.
When the chosen estimator of $\gamma^2$ has high variance, under heavy
contamination or small samples, the sandwich may substantially underestimate
$\mathrm{Var}(\hat{\boldsymbol\beta})$.
Both limitations are resolved by the bootstrap.

\subsection[Bootstrap inference]{Bootstrap Inference} \label{subsec:bootstrap}

\subsubsection[Motivation]{Motivation}

The sandwich treats $\hat{\mathbf{K}}$ as fixed, but $k_{ii} =
G(y_i,\mathbf{x}_i^\top\hat{\boldsymbol\beta})$ depends on
$\hat{\boldsymbol\beta}$, which depends on $\hat\gamma^2$.
The pairs bootstrap \citep{Efron:1979,Efron+Tibshirani:1993} re-executes the
entire GKRReg algorithm (including re-estimating $\gamma^2$) on each
resampled dataset, capturing all sources of variability.

\subsubsection[Algorithm]{Algorithm}

Let $\mathcal{D} = \{(\mathbf{x}_i,y_i)\}_{i=1}^n$.
For $b = 1,\ldots,B$:
\begin{enumerate}
  \item Draw $\mathcal{D}^*_b$ by sampling $n$ pairs with replacement
        from $\mathcal{D}$.
  \item Re-estimate $\hat\gamma^{2*}_b$ from $\mathcal{D}^*_b$ using the
        same estimator (S1--S4 or \code{"auto"}) as in the original fit.
  \item Run GKRReg on $\mathcal{D}^*_b$ with $\hat\gamma^{2*}_b$ until
        convergence, obtaining $\hat{\boldsymbol\beta}^*_b$.
\end{enumerate}
Non-convergent replicates are discarded; the effective count
$B_{\mathrm{ok}}\le B$ is reported.

\subsubsection[Confidence intervals]{Confidence Intervals}

We propose three different approaches for obtaining confidence intervals for the vector of parameters $\bm{\beta}$. The percentile interval uses the empirical quantiles of the bootstrap
distribution. The normal interval, given by
$\hat\beta_j \pm z_{\alpha/2}\widehat{\mathrm{se}}^*_j$.
Finally, the \emph{bias-corrected and accelerated} (BCa) interval \citep{Efron:1987}
corrects for bias and skewness via a bias-correction term
$\hat z_0 = \Phi^{-1}\!\left(B^{-1}\#\{\hat\beta^*_{b,j}<\hat\beta_j\}\right)$
and an acceleration $\hat a$ estimated by jackknife \citep{Efron:1982};
BCa typically achieves better coverage in small and moderate samples
\citep{DiCiccio+Efron:1996} and is the recommended default.

\subsubsection[Bootstrap p-values]{Bootstrap $p$-Values}

To test $H_0:\beta_j=0$ we use the centred bootstrap-$t$ approach
\citep{Davison+Hinkley:1997,Hall:1992}, The two-sided $p$-value is given by
\begin{equation}
  p_j = \frac{1}{B}\,
        \#\!\left\{b :
          |\hat\beta^*_{b,j}-\hat\beta_j| \ge |\hat\beta_j|
        \right\},
  \label{eq:pval}
\end{equation}
bounded below by $1/B$.
No additional computation beyond the bootstrap replicates is required.

\subsection[Choosing between the two approaches]{Choosing between the two approaches} \label{subsec:choice}

When both methods are computed, \pkg{gkrreg} automatically issues a warning whenever the relative discrepancy between the sandwich and bootstrap standard errors exceeds a threshold $\tau$ for any coefficient $j$, that is, when
\begin{equation*}
  \frac{|\widehat{\mathrm{se}}_{\mathrm{sw},j} - \widehat{\mathrm{se}}^*_j|}
       {\widehat{\mathrm{se}}^*_j} > \tau,
\end{equation*}
with default $\tau = 0.25$. This divergence is symptomatic of cases where the sandwich assumption of negligible $\hat\gamma^2$ variability does not hold. Because the sandwich estimator treats $\hat\gamma^2$ as fixed, its standard errors will be systematically smaller than their bootstrap counterparts whenever the estimation of $\gamma^2$ contributes substantially to the overall uncertainty in $\hat{\boldsymbol\beta}$, as typically occurs under heavy contamination or small sample sizes. On the other hand, a large discrepancy signals that the analytic approximation is
insufficient and that bootstrap inference should be preferred.

Additionally, when only the sandwich is available, \pkg{gkrreg} proactively emits a note if either of two conditions is met, i.e.,  the sample size is small ($n < 50$), or more than 10\% of observations have received a kernel weight below~$0.01$. The latter criterion identifies observations that the algorithm has effectively
discarded. For example, a kernel weight of $k_{ii} < 0.01$ corresponds to a residual
exceeding approximately $2.15\sqrt{\hat\gamma^2}$, which is a genuinely large deviation regardless of the scale of $\hat\gamma^2$. Moreover, a contamination rate above 10\% is precisely the scenario where the variability of $\hat\gamma^2$ is highest and the sandwich approximation least reliable. Both conditions can be checked automatically at negligible computational cost, providing the user with actionable guidance without requiring the bootstrap to have been run. Table~\ref{tab:inference} summarises the main characteristics of the two
inference procedures.
\begin{table}[h!]
\centering
\caption{\label{tab:inference}Comparison of sandwich (HC0) and bootstrap
  inference for GKRReg. $B$ = number of bootstrap replicates.}
\begin{tabular}{lll}
\toprule
Property & Sandwich (HC0) & Bootstrap (BCa) \\
\midrule
Computational cost
  & $O(np^2)$ & $O(Bnp^2)$ \\
Accounts for $\hat\gamma^2$ variability
  & No & Yes \\
Finite-sample correction
  & No (HC0) & Implicit via resampling \\
Reliable for small $n$
  & Limited & Yes \\
Corrects for skewness
  & No & Yes \\
Deterministic
  & Yes & No (seed-based) \\
Recommended scenario
  & $n\ge 50$, mild contamination
  & $n<50$ or heavy contamination \\
\bottomrule
\end{tabular}
\end{table}
%


\section[The gkrreg package]{The \pkg{gkrreg} Package} \label{sec:package}

\subsection[Overview]{Overview}

The \pkg{gkrreg} package \citep{gkrreg:2025} is available on CRAN and
implements all the methods described in Sections~\ref{sec:method} and
\ref{sec:inference}.
Six real datasets from the robust regression literature are bundled
(Table~\ref{tab:datasets}).
\begin{table}[t!]
\centering
\caption{\label{tab:datasets}Datasets bundled with \pkg{gkrreg}.}
\begin{tabular}{llrrl}
\toprule
Dataset & Response & $n$ & $p$ & Outlier type \\
\midrule
\code{belgium\_calls} & International calls & 24 & 1 & Y-space \\
\code{cloud\_point}   & Cloud point temp.   & 19 & 1 & Leverage \\
\code{kootenay}       & River flow          & 13 & 1 & X-space \\
\code{delivery}       & Delivery time       & 25 & 2 & Bad leverage \\
\code{mammals}        & log(brain mass)     & 62 & 1 & Leverage (log scale) \\
\code{stars\_cyg}     & log(luminosity)     & 47 & 1 & Bad leverage \\
\bottomrule
\end{tabular}
\end{table}

\subsection[Model fitting]{Model Fitting: \fct{gkrr}}
\label{subsec:gkrr_func}

The main function follows the standard \proglang{R} formula interface:
\begin{Code}
gkrr(formula, data, sigma_method = c("s1", "s2", "s3", "s4", "auto"),
     alpha = 0.5, tol = 1e-10, maxit = 100,
     boot = FALSE, boot_args = list(), auto_args = list(), conf = 0.95)
\end{Code}
The \code{sigma\_method} argument accepts \code{"s1"}--\code{"s4"} or
\code{"auto"} (Section~\ref{subsec:gamma}), or a positive numeric value
that fixes $\gamma^2$ directly.
The \code{s4} option requires the \pkg{sm} package \citep{Bowman+Azzalini:1997}.
When \code{"auto"} is requested, a brief bootstrap with $B=99$ replicates
(configurable via \code{auto\_args = list(B = ...)}) is run to select among
S1--S4 by out-of-bag MSE; a message announces the selected method and the
comparative OOB MSEs.

\subsubsection{Automatic $\gamma^2$ Selection}
\label{subsec:auto}

For $b = 1,\ldots,B_{\mathrm{auto}}$, the algorithm draws a bootstrap
replicate $\mathcal{D}^*_b$, fits GKRReg with each candidate estimator
S1--S4, and computes the OOB mean squared prediction error on observations
not selected in $\mathcal{D}^*_b$.
The estimator with the lowest average OOB MSE is selected.
This approach adds computational cost but requires no prior knowledge about
the outlier scenario.

\subsection[Inference]{Inference: \fct{summary} and \fct{gkrr\_boot}}
\label{subsec:inference_func}

The sandwich covariance matrix~\eqref{eq:sandwich} is computed automatically during model fitting and is always available via \fct{vcov}. The \fct{summary} function displays a coefficient table analogous to \fct{summary.lm}. We illustrate the inference capabilities of \pkg{gkrreg} on the \code{mammals} dataset~\citep{Allison+Cicchetti:1976}, which contains body mass ($kg$) and brain mass ($g$) for 62 mammal species. On the natural log scale, the relationship is approximately linear but the African and Asian elephants are high-leverage observations. We create the log-transformed variables and fit GKRReg with the default sandwich inference:
\begin{CodeChunk}
\begin{CodeInput}
R> library("gkrreg")
R> data("mammals")
R> mammals$log_body  <- log(mammals$body_mass)
R> mammals$log_brain <- log(mammals$brain_mass)
R> fit <- gkrr(log_brain ~ log_body, data = mammals, sigma_method = "s3")
R> summary(fit)
\end{CodeInput}
\begin{CodeOutput}
Call:
gkrr(formula = log_brain ~ log_body, data = mammals, sigma_method = "s3")

Residuals:
    Min      1Q  Median      3Q     Max
-1.5832 -0.3770  0.0661  0.5718  2.0946

Coefficients:
            Estimate Std. Error CI 95% lower CI 95% upper   p-value
(Intercept)   2.0086     0.1164       1.7805       2.2367 2.000e-16 ***
log_body      0.7468     0.0186       0.7104       0.7832 2.000e-16 ***
---
Signif. codes:  0 '***' 0.001 '**' 0.01 '*' 0.05 '.' 0.1 ' ' 1
(Sandwich SE (asymptotic Wald z-test))

gamma^2: 0.482  |  method: s3  |  iterations: 26  |  converged: TRUE
R-squared: 0.9178  |  Weighted R-squared: 0.9775
Kernel weights -- min: 0.0001  mean: 0.6003  max: 1.0000
\end{CodeOutput}
\end{CodeChunk}
The sandwich $p$-values confirm a highly significant positive relationship between log body mass and log brain mass ($p < 2 \times 10^{-16}$), with an elasticity estimate of $\hat\beta_1 = 0.747$. The full sandwich covariance matrix is accessible via \fct{vcov}:
\begin{CodeChunk}
\begin{CodeInput}
R> round(vcov(fit), 6)
\end{CodeInput}
\begin{CodeOutput}
            (Intercept)  log_body
(Intercept)    0.013547 -0.000551
log_body      -0.000551  0.000345
\end{CodeOutput}
\end{CodeChunk}
The negative off-diagonal element reflects the typical negative correlation between intercept and slope estimates in simple regression.

A pre-computed \code{gkrr\_boot} object can be passed directly to
\fct{summary}, providing bootstrap inference without redundant computation:
\begin{CodeChunk}
\begin{CodeInput}
R> boot <- gkrr_boot(fit, B = 999, type = "bca", seed = 1)
R> summary(fit, boot = boot)
\end{CodeInput}
\begin{CodeOutput}
Call:
gkrr(formula = log_brain ~ log_body, data = mammals, sigma_method = "s3")

Residuals:
    Min      1Q  Median      3Q     Max
-1.5832 -0.3770  0.0661  0.5718  2.0946

Coefficients:
            Estimate Std. Error CI 95% lower CI 95% upper p-value
(Intercept)   2.0086     0.1389       1.7846       2.3665   0.001 **
log_body      0.7468     0.0273       0.6770       0.7919   0.001 **
---
Signif. codes:  0 '***' 0.001 '**' 0.01 '*' 0.05 '.' 0.1 ' ' 1
(Bootstrap p-values: centred-t, B = 998, BCA CI)

gamma^2: 0.482  |  method: s3  |  iterations: 26  |  converged: TRUE
R-squared: 0.9178  |  Weighted R-squared: 0.9775
Kernel weights -- min: 0.0001  mean: 0.6003  max: 1.0000
Warning message:
Sandwich and bootstrap SEs diverge by more than 25% for: log_body [32%, sand
wich < bootstrap (sandwich may underestimate SE)].
  This may indicate small n, heavy outlier contamination, or high
  variability in the gamma^2 estimator. Bootstrap inference is
  recommended in this case.
\end{CodeOutput}
\end{CodeChunk}
Both inference methods agree. The scaling relationship is highly significant regardless of whether sandwich or bootstrap standard errors are used. The bootstrap standard errors are moderately larger, capturing the additional uncertainty from re-estimating $\hat\gamma^2$ on each replicate.
%

\textbf{Note on the reported R-squared.}
The classical coefficient of determination $R^2 = 1 - SS_{\text{res}}/SS_{\text{tot}}$
is guaranteed to lie in $[0,1]$ for OLS because OLS minimises $SS_{\text{res}}$ ensuring the fitted model is never worse than the intercept-only
model $\hat\mu = \bar y$.
This guarantee does not extend to other estimators.
For GKRReg, the algorithm intentionally assigns near-zero kernel weights to
outliers and/or leverage points, producing large residuals for those observations.
When outlier contamination is considerable, the resulting $SS_{\text{res}}$ can
exceed $SS_{\text{tot}}$, yielding $R^2 < 0$.
A negative $R^2$ therefore does not indicate a poor fit; it reflects that the
model is fitting the majority of the data well while deliberately ignoring the
anomalous observations.

The \emph{weighted} $R^2$ reported alongside the classical value,
\begin{equation*}
  R^2_w = 1 - \frac{\displaystyle\sum_{i=1}^n k_{ii}(y_i - \hat\mu_i)^2}
                   {\displaystyle\sum_{i=1}^n k_{ii}(y_i - \bar y_w)^2},
  \qquad
  \bar y_w = \frac{\sum_{i=1}^n k_{ii} y_i}{\sum_{i=1}^n k_{ii}},
\end{equation*}
down-weights outliers in proportion to their kernel weight $k_{ii}$, providing
a measure of goodness-of-fit restricted to the observations the model considers
informative.
For the \code{mammals} example, both $R^2 = 0.918$ and $R^2_w = 0.978$
are positive and high, reflecting a strong fit on the non-outlying species.
In datasets with severe contamination, such as \code{belgium\_calls}
(Section~\ref{subsec:belgium}), the classical $R^2$ may be negative while
$R^2_w$ remains informative.

\subsection[Diagnostic analysis]{Diagnostic Analysis: \fct{plot.gkrr} and \fct{plot.gkrr\_boot}} \label{subsec:plots}

Function \fct{plot.gkrr} provides six panels (\code{which = 1:6}) allowing
to perform a complete diagnostic analysis for the model.
An important aspect of this function is that point size is inversely
proportional to $k_{ii}$. Observations with small weights appear large and
red, while well-fitted observations appear small and blue.
Panel~3 (kernel weight versus residual) overlays the theoretical curve
$G(e) = \exp(-e^2/\hat\gamma^2)$, making the down-weighting mechanism
directly visible.
Function \fct{plot.gkrr\_boot} provides two panels. The first is a histogram of bootstrap
replicates with the shaded BCa interval (panel~1), and the second is a scatterplot matrix of
all pairs of bootstrap replicates with 95\% confidence ellipses (panel~2),
revealing between-coefficient dependencies that marginal intervals do not
capture.

The following code illustrates these diagnostic tools on the \code{mammals}
fit from Section~\ref{subsec:inference_func}:
\begin{CodeChunk}
\begin{CodeInput}
R> par(mfrow = c(1, 2))
R> plot(fit, which = 3, ask = FALSE)
R> plot(fit, which = 4, ask = FALSE)
R> par(mfrow = c(1, 1))
\end{CodeInput}
\end{CodeChunk}
\begin{figure}[ht!]
\centering
\includegraphics[width=\textwidth]{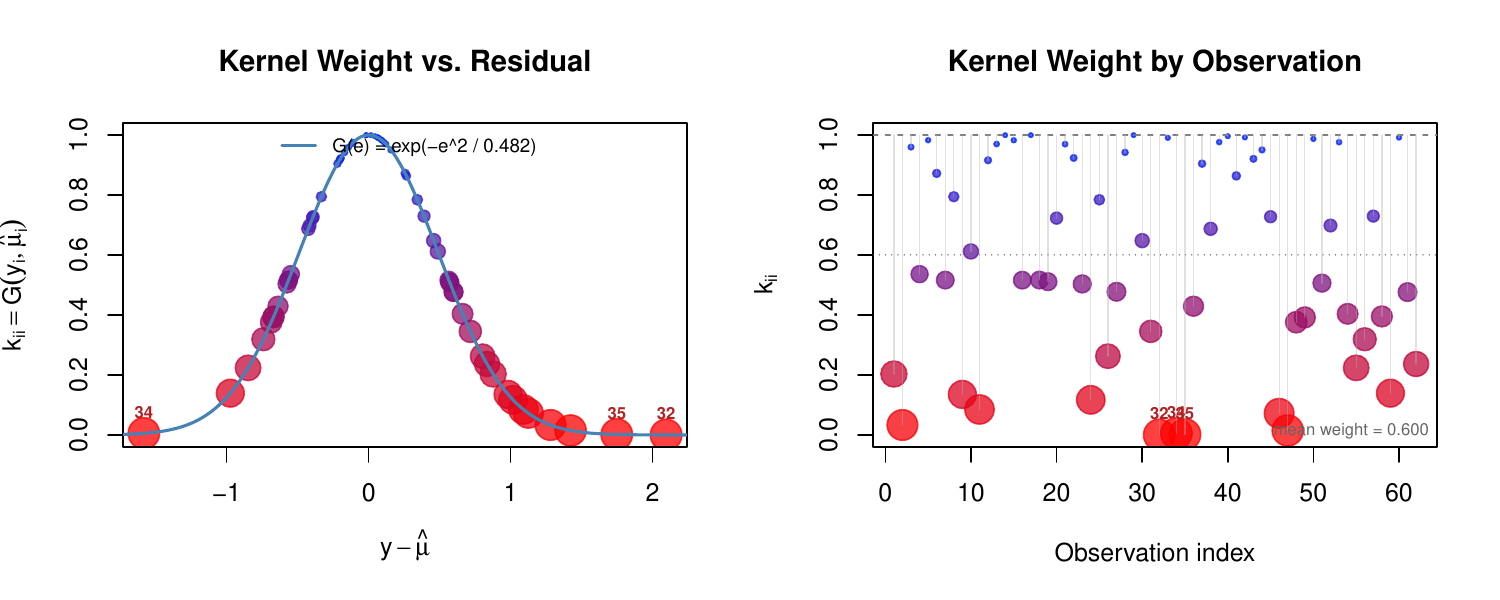}
\caption{\label{fig:mammals_panels}Diagnostic panels for the \code{mammals}
  fit.  \emph{Left}: kernel weight vs.\ residual with the theoretical curve
  $G(e)=\exp(-e^2/\hat\gamma^2)$ overlaid.
  \emph{Right}: kernel weight vs.\ observation index, identifying the
  down-weighted species.}
\end{figure}
Two diagnostic plots for the \code{mammals} fit are shown in~Figure~\ref{fig:mammals_panels}. The left panel overlays the theoretical kernel curve on the weight--residual scatter. Observations with large residuals, most notably the elephant species and a handful of other large mammals, fall in the tails of the curve and receive near-zero weights. The right panel identifies these observations by index, showing that the down-weighting is concentrated among a specific subset of species rather than spread uniformly across the dataset. The remaining 62 species are fitted with high fidelity, as reflected by the weighted $R^2_w = 0.978$. Figure \ref{fig:mammals_boot_hist} illustrates the bootstrap histogram for the coefficients estimates on \code{mammals} fit. The shaded region indicates the 95\% BCa confidence intervals with the estimated coefficients centred well away from zero, confirming the significance of the parameter estimates.

\begin{CodeChunk}
\begin{CodeInput}
R> par(mfrow = c(2,1))
R> plot(boot, which = 1, ask = FALSE)
R> par(mfrow = c(1,1))
\end{CodeInput}
\end{CodeChunk}
\begin{figure}[h!]
\centering
\includegraphics[width=\textwidth]{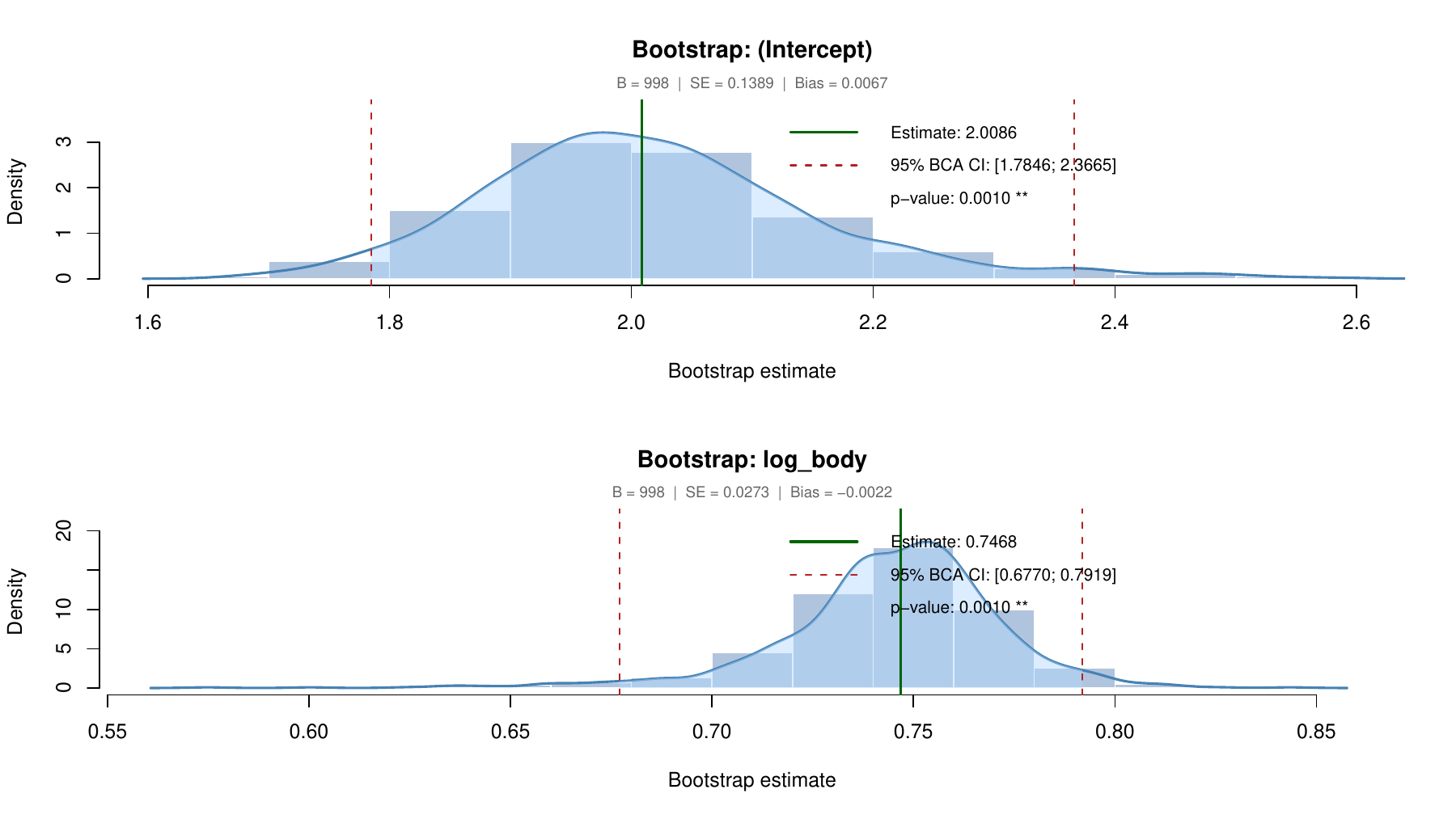}
\caption{\label{fig:mammals_boot_hist}Bootstrap histogram for the coefficients estimates on \code{mammals} fit ($B = 999$ BCa replicates).}
\end{figure}


\section[Illustrations]{Illustrations} \label{sec:illustrations}

We illustrate \pkg{gkrreg} on the six bundled datasets, each highlighting a different functionality. Except in Section~\ref{subsec:stars}, where a direct comparison with OLS is the focus, OLS fits are not shown in order to keep the emphasis on the GKRReg-specific inference and diagnostic tools.

\subsection[Belgium international calls: Y-space outliers]{Belgium International Calls: Y-Space Outliers}
\label{subsec:belgium}

The \code{belgium\_calls} dataset \citep{Rousseeuw+Leroy:1987} records
international telephone calls from Belgium (1950--1973).
Observations~15--20 (years 1964--1969) are Y-space outliers caused by a
recording error: calls were measured in total minutes rather than number of calls. We fit GKRReg with \code{sigma\_method = "s3"} (recommended for heavy Y-space contamination) and focus on the diagnostic panels, which make the down-weighting mechanism directly visible.
\begin{CodeChunk}
\begin{CodeInput}
R> data("belgium_calls")
R> fit_bel <- gkrr(calls ~ year, data = belgium_calls, sigma_method = "s3")
R> summary(fit_bel)
\end{CodeInput}
%
\begin{CodeOutput}
Call:
gkrr(formula = calls ~ year, data = belgium_calls, sigma_method = "s3")

Residuals:
    Min      1Q  Median      3Q     Max
-0.6165 -0.1917 -0.0271  3.5213 18.4537

Coefficients:
            Estimate Std. Error CI 95% lower CI 95% upper   p-value
(Intercept)  -6.5764     1.1145      -8.7609      -4.3920 3.621e-09 ***
year          0.1351     0.0203       0.0954       0.1748 2.529e-11 ***
---
Signif. codes:  0 '***' 0.001 '**' 0.01 '*' 0.05 '.' 0.1 ' ' 1
(Sandwich SE (asymptotic Wald z-test))

gamma^2: 31.61  |  method: s3  |  iterations: 9  |  converged: TRUE
R-squared: -0.1219  |  Weighted R-squared: 0.5499
Kernel weights -- min: 0.0000  mean: 0.7501  max: 1.0000
Note: sandwich inference may be less reliable here (n = 24 (small sample);
12% of observations received near-zero kernel weight (< 0.01)).
  Consider bootstrap inference via boot = TRUE in gkrr() or
  summary(fit, boot = gkrr_boot(fit)).
\end{CodeOutput}
\end{CodeChunk}
The sandwich $p$-values confirm a highly significant positive trend
($p < 10^{-8}$) after correcting for the outliers.
The classical $R^2 = -0.12$ is negative because the six erroneous
observations have large residuals by design. On the other hand, the weighted $R^2_w = 0.55$
reflects the goodness of fit on the remaining 18 observations
(see the note in Section~\ref{subsec:inference_func}).
\begin{CodeChunk}
\begin{CodeInput}
R> par(mfrow = c(1, 2))
R> plot(fit_bel, which = 3, ask = FALSE)
R> plot(fit_bel, which = 4, ask = FALSE)
R> par(mfrow = c(1, 1))
\end{CodeInput}
\end{CodeChunk}
\begin{figure}[h!]
\centering
\includegraphics[width=\textwidth]{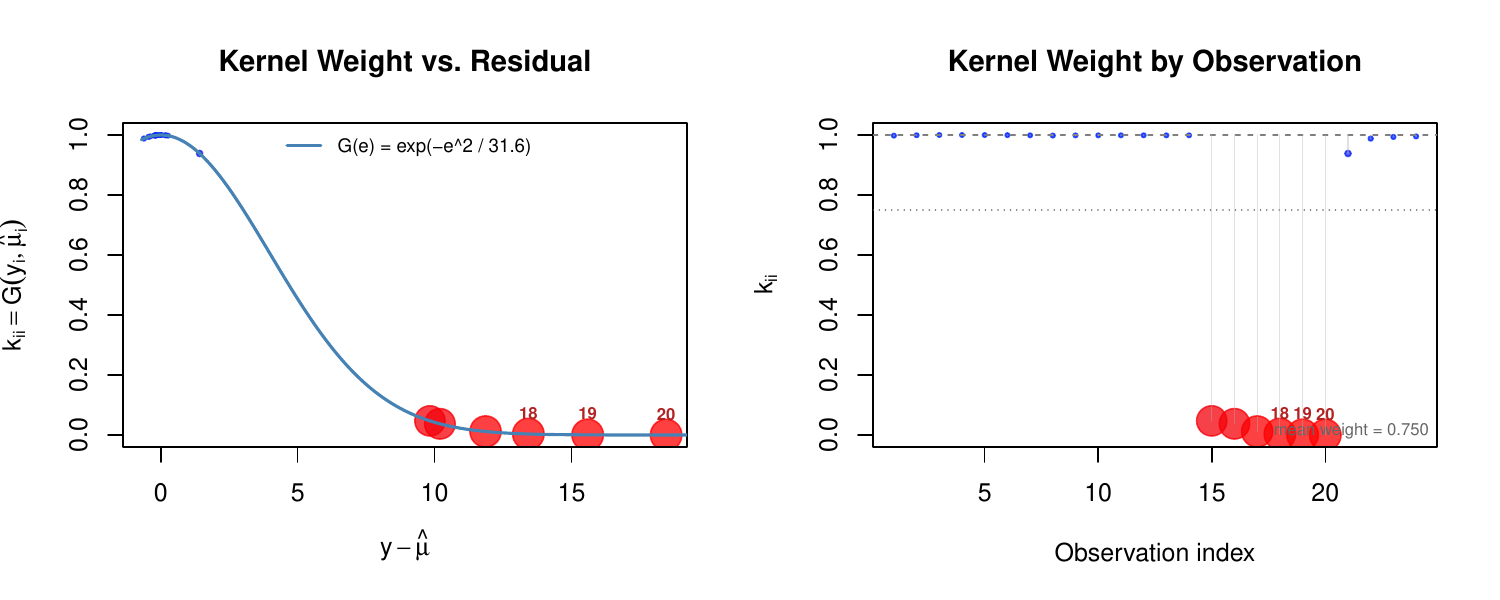}
\caption{\label{fig:belgium_panels}Diagnostic panels for the
  \code{belgium\_calls} fit.  \emph{Left}: kernel weight vs.\ residual with
  the theoretical curve $G(e)=\exp(-e^2/\hat\gamma^2)$ overlaid. \emph{Right}: kernel weight vs.\ observation index, identifying the six
  erroneous observations unambiguously.}
\end{figure}

Figure~\ref{fig:belgium_panels} illustrates two diagnostic plots. On the right, GKRReg assigns near-zero weights to observations 15--20 (visible in Figure 1, right), recovering the true trend. On the other hands, the OLS will be severely
distorted by these six outlier observations.

\subsection[Delivery time: inference in multiple regression]{Delivery Time: Inference in Multiple Regression}
\label{subsec:delivery}

\begin{figure}[h!]
\centering
\includegraphics[width=\textwidth]{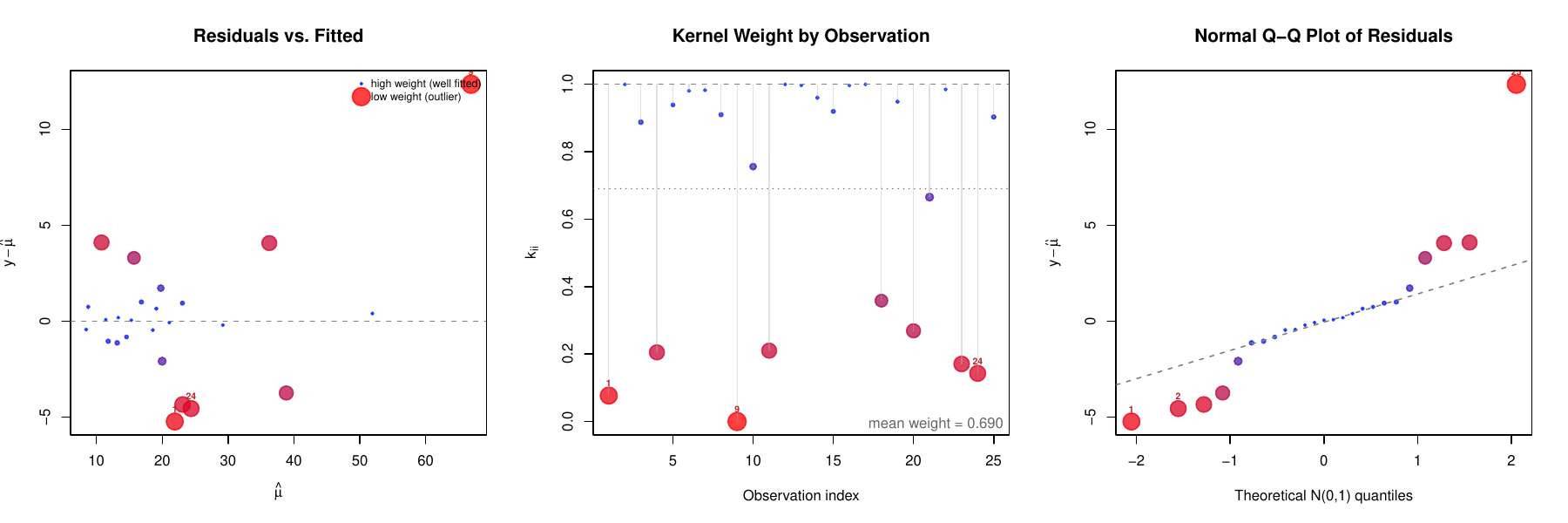}
\caption{\label{fig:delivery_diag}Diagnostic panels for the \code{delivery}
  fit.  \emph{Left}: residuals vs.\ fitted values.
  \emph{Centre}: kernel weight vs.\ observation index.
  \emph{Right}: QQ-plot of residuals coloured by kernel weight.}
\end{figure}

The \code{delivery} dataset~\citep{Montgomery+Peck:1992} contains 25
observations on soft-drink delivery time with two predictors:
\code{n\_products} (number of cases delivered) and \code{distance}
(distance walked, in feet).
Observation~9 (\code{n\_products = 30}, \code{distance = 1460}) is a bad
leverage point far outside the bulk of the predictor space.
We use \code{sigma\_method = "s3"} and focus on the full coefficient table
with sandwich inference and on the diagnostic panels most informative for
leverage detection.
\begin{CodeChunk}
\begin{CodeInput}
R> data("delivery")
R> fit_del <- gkrr(delivery_time ~ n_products + distance,
+                  data = delivery, sigma_method = "s3")
R> summary(fit_del)
\end{CodeInput}
\begin{CodeOutput}
Call:
gkrr(formula = delivery_time ~ n_products + distance,
    data = delivery, sigma_method = "s3")

Residuals:
     Min       1Q   Median       3Q      Max
-5.2212  -1.0412   0.0559   0.9440  12.3448

Coefficients:
            Estimate Std. Error CI 95% lower CI 95% upper   p-value
(Intercept)   4.0607     0.5908       2.9027       5.2187 6.298e-12 ***
n_products    1.3891     0.0883       1.2161       1.5621 2.000e-16 ***
distance      0.0145     0.0034       0.0078       0.0212 2.238e-05 ***
---
Signif. codes:  0 '***' 0.001 '**' 0.01 '*' 0.05 '.' 0.1 ' ' 1
(Sandwich SE (asymptotic Wald z-test))

gamma^2: 10.62  |  method: s3  |  iterations: 23  |  converged: TRUE
R-squared: 0.9496  |  Weighted R-squared: 0.9814
Kernel weights -- min: 0.0000  mean: 0.6904  max: 0.9997
Note: sandwich inference may be less reliable here (n = 25 (small sample)).
  Consider bootstrap inference via boot = TRUE in gkrr() or
  summary(fit, boot = gkrr_boot(fit)).
\end{CodeOutput}
\end{CodeChunk}
Both predictors are highly significant ($p < 10^{-4}$).
The positive $R^2 = 0.95$ and weighted $R^2_w = 0.98$ confirm an excellent
fit on the non-outlying observations. The diagnostic panels confirm which observation is responsible for the down-weighting and assess the normality of the residuals after the robust fitting. Note in Figure~\ref{fig:delivery_diag} (center) that observation~9
receives near-zero weight, confirming its status as a bad leverage point. Moreover, in the right panel observation~9 appears as a large red point far from the reference line.
\begin{CodeChunk}
\begin{CodeInput}
R> par(mfrow = c(1, 3))
R> plot(fit_del, which = 1, ask = FALSE)
R> plot(fit_del, which = 4, ask = FALSE)
R> plot(fit_del, which = 5, ask = FALSE)
R> par(mfrow = c(1, 1))
\end{CodeInput}
\end{CodeChunk}

\subsection[Kootenay river flow: bootstrap inference for small samples]{%
             Kootenay River Flow: Bootstrap Inference for Small Samples}
\label{subsec:kootenay}

\begin{figure}[h!]
\centering
\includegraphics[width=\textwidth]{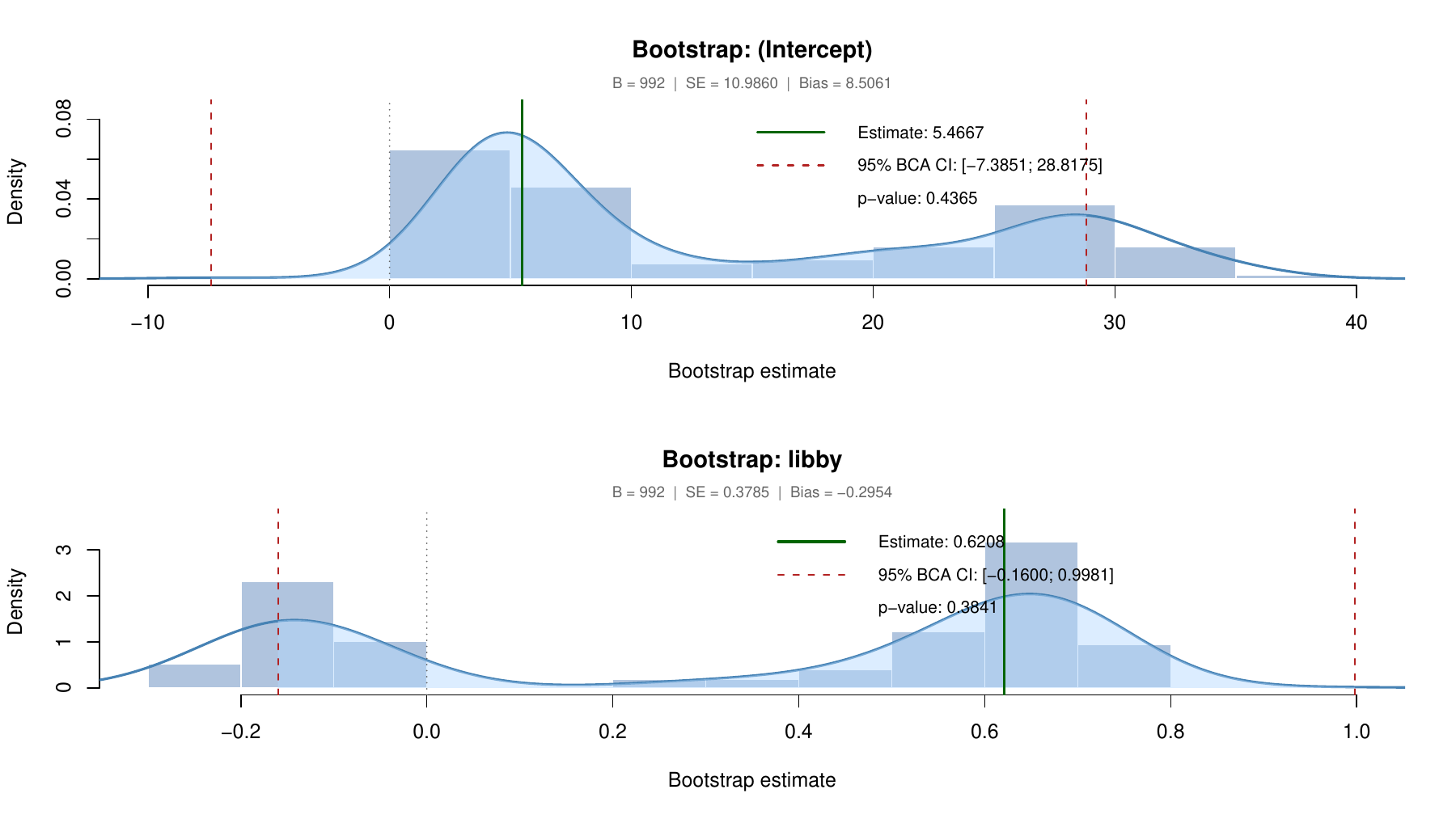}
\caption{\label{fig:kootenay_boot}Bootstrap distribution of the
  \code{kootenay} fit coefficients ($B = 999$ BCa replicates).}
\end{figure}

The \code{kootenay} dataset \citep{Neter+:1996} contains annual water-flow
measurements at two gauging stations on the Kootenay river (Libby and
Newgate, Montana/British Columbia) for the years 1931 to 1943.
The 1934 observation exhibits an anomalously large Libby measurement (77.6 vs.\
a typical range of 17--39), making it an X-space outlier.
With $n = 13$ observations, the sandwich estimator is unreliable and bootstrap inference is strongly recommended; we therefore adopt estimator S1, which is specifically suited to handle X-space contamination.
\begin{CodeChunk}
\begin{CodeInput}
R> data("kootenay")
R> fit_koot  <- gkrr(newgate ~ libby, data = kootenay, sigma_method = "s1")
R> boot_koot <- gkrr_boot(fit_koot, B = 999, type = "bca", seed = 1)
R> summary(fit_koot, boot = boot_koot)
\end{CodeInput}
\begin{CodeOutput}
Call:
gkrr(formula = newgate ~ libby, data = kootenay, sigma_method = "s1")

Residuals:
     Min       1Q   Median       3Q      Max 
-37.9443  -0.8063  -0.1029   1.0230   1.8066 

Coefficients:
            Estimate Std. Error CI 95% lower CI 95% upper p-value  
(Intercept)   5.4667    10.9860      -7.3851      28.8175  0.4365  
libby         0.6208     0.3785      -0.1600       0.9981  0.3841  
---
Signif. codes:  0 '***' 0.001 '**' 0.01 '*' 0.05 '.' 0.1 ' ' 1
(Bootstrap p-values: centred-t, B = 992, BCA CI)

gamma^2: 34.05  |  method: s1  |  iterations: 11  |  converged: TRUE
R-squared: -4.5241  |  Weighted R-squared: 0.9010
Kernel weights -- min: 0.0000  mean: 0.8752  max: 0.9997
Warning message:
Sandwich and bootstrap SEs diverge by more than 25% for: (Intercept) [84%,
sandwich < bootstrap (sandwich may underestimate SE)]; libby [83%, sandwich
< bootstrap (sandwich may underestimate SE)].
  This may indicate small n, heavy outlier contamination, or high
  variability in the gamma^2 estimator. Bootstrap inference is
  recommended in this case.
\end{CodeOutput}
\end{CodeChunk}
The 1934 observation receives zero weight (\code{min: 0.0000}), and the
weighted $R^2_w = 0.90$ indicates a strong linear relationship between the
two gauging stations once the anomalous year is discounted.
The non-significant bootstrap $p$-values for this small dataset reflect the genuine uncertainty in the slope estimate when only 12 effective observations are available. The divergence warning issued by \fct{summary} confirms that bootstrap inference is the appropriate choice here. Figure~\ref{fig:kootenay_boot} illustrates the bootstrap histogram for the intercept in the \code{kootney} fit. The wide shaded BCa intervals includes the zero, confirming the non-significance of the parameter estimates.
\begin{CodeChunk}
\begin{CodeInput}
R> par(mfrow = c(2,1))
R> plot(boot_koot, which = 1, ask = FALSE)
R> par(mfrow = c(1,1))
\end{CodeInput}
\end{CodeChunk}

\subsection[Cloud point: automatic $gamma^2$ selection]{Cloud Point: Automatic $\gamma^2$ Selection} \label{subsec:cloud}

\begin{figure}[h!]
\centering
\includegraphics[width=\textwidth]{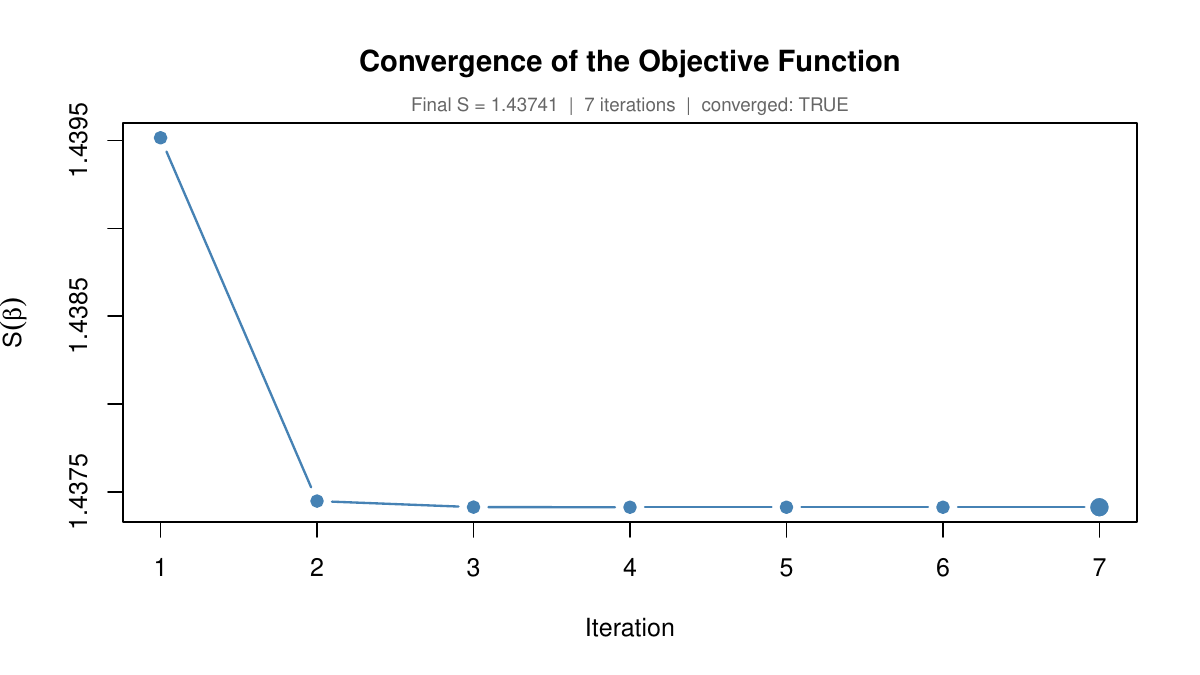}
\caption{\label{fig:cloud_diag}Convergence of the objective function $S(\boldsymbol\beta)$ for the \code{cloud\_point} fit with \code{sigma\_method = "auto"}.}
\end{figure}

The \code{cloud\_point} dataset \citep{Draper+Smith:1998} measures the cloud point temperature ($^\circ$C) of a liquid mixture of isomers as a function of \code{percentage\_i8} (percentage of one isomer). Three observations at \code{percentage\_i8 = 0} act as leverage points.
We illustrate \code{sigma\_method = "auto"}, which selects the optimal
estimator by out-of-bag bootstrap MSE without requiring the user to specify the contamination scenario.
\begin{CodeChunk}
\begin{CodeInput}
R> data("cloud_point")
R> fit_auto <- gkrr(cloud_point ~ percentage_i8, data = cloud_point,
+                   sigma_method = "auto", auto_args = list(B = 99, seed = 1))
\end{CodeInput}
\begin{CodeOutput}
sigma_method = "auto": running bootstrap selection (B = 99 replicates).
sigma_method = "auto": "s2" selected (avg. OOB MSE: s1=1.1627, s2=0.7847,
s3=1.1574, s4=0.9520)
\end{CodeOutput}
\end{CodeChunk}
\begin{CodeChunk}
\begin{CodeInput}
R> summary(fit_auto)
\end{CodeInput}
\begin{CodeOutput}
Call:
gkrr(formula = cloud_point ~ percentage_i8, data = cloud_point, 
    sigma_method = "auto", auto_args = list(B = 99, seed = 1))

Residuals:
    Min      1Q  Median      3Q     Max 
-1.4970 -0.4489  0.1752  0.5770  0.9511 

Coefficients:
              Estimate Std. Error CI 95% lower CI 95% upper   p-value    
(Intercept)     23.397     0.4035      22.6062      24.1878 2.000e-16 ***
percentage_i8    1.038     0.0677       0.9052       1.1707 2.000e-16 ***
---
Signif. codes:  0 '***' 0.001 '**' 0.01 '*' 0.05 '.' 0.1 ' ' 1
(Sandwich SE (asymptotic Wald z-test))

gamma^2: 12.29  |  method: auto(s2)  |  iterations: 7  |  converged: TRUE
R-squared: 0.9547  |  Weighted R-squared: 0.9557
Kernel weights -- min: 0.8333  mean: 0.9622  max: 0.9997
Note: sandwich inference may be less reliable here (n = 19 (small sample)).
  Consider bootstrap inference via boot = TRUE in gkrr() or
  summary(fit, boot = gkrr_boot(fit)).
\end{CodeOutput}
\end{CodeChunk}
The \code{"auto"} procedure selected S2 (pairwise median), achieving the lowest OOB MSE (0.7847), consistent with the moderate leverage-type contamination present in this dataset.
The \code{method: auto(s2)} label in the output provides full traceability of the selection. The code below illustrates Figure~\ref{fig:cloud_diag}. The panel shows the convergence of the objective function $S(\boldsymbol\beta)$ over the 7 IRWLS iterations.
\begin{CodeChunk}
\begin{CodeInput}
R> plot(fit_auto, which = 6, ask = FALSE)
\end{CodeInput}
\end{CodeChunk}

\subsection[Mammals: bootstrap scatter-plot matrix]{Mammals: Bootstrap Scatter-Plot Matrix} \label{subsec:mammals}

\begin{figure}[h!]
\centering
\includegraphics[width=0.65\textwidth]{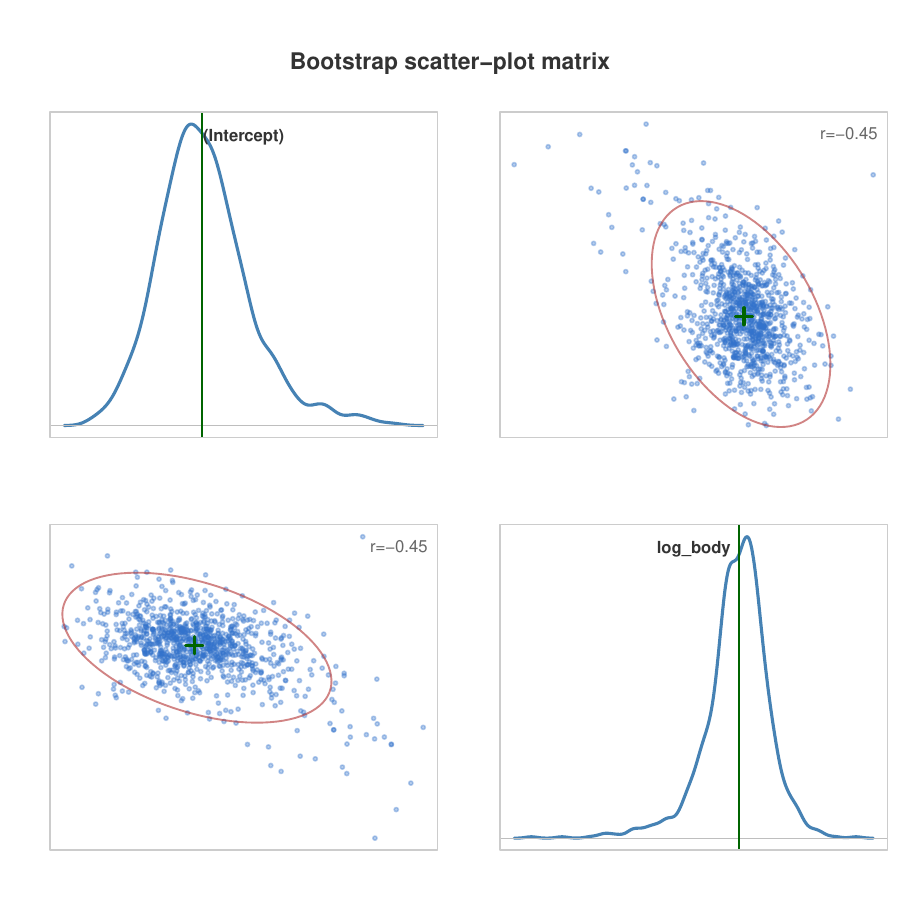}
\caption{\label{fig:mammals_boot_scatter}Bootstrap scatter-plot matrix for
  the \code{mammals} fit ($B = 999$ BCa replicates).}
\end{figure}

The \code{mammals} dataset was used in Section~\ref{sec:package} to illustrate the core inference workflow. The code below focuses on the bootstrap scatter-plot matrix (panel~2 of \fct{plot.gkrr\_boot}), which reveals the joint distribution of the bootstrap replicates and complements the marginal intervals reported by \fct{summary}. Figure~\ref{fig:mammals_boot_scatter} illustrates a negative correlation between intercept and slope bootstrap replicates. This joint dependence, which is typical of simple linear regression, is also captured by the asymptotic correlation derived from \fct{vcov} (approximately $-0.26$). The confidence ellipse confirms that the two parameters are jointly well-determined.
\begin{CodeChunk}
\begin{CodeInput}
R> data("mammals")
R> mammals$log_body  <- log(mammals$body_mass)
R> mammals$log_brain <- log(mammals$brain_mass)
R> fit_mam  <- gkrr(log_brain ~ log_body, data = mammals, sigma_method = "s3")
R> boot_mam <- gkrr_boot(fit_mam, B = 999, type = "bca", seed = 1)
R> plot(boot_mam, which = 2, ask = FALSE)
\end{CodeInput}
\end{CodeChunk}

\subsection[Hertzsprung-Russell diagram: kernel weight diagnostic and comparison with OLS]{Hertzsprung-Russell Diagram: Kernel Weight Diagnostic and Comparison with OLS}
\label{subsec:stars}

\begin{figure}[h!]
\centering
\includegraphics[width=0.6\textwidth]{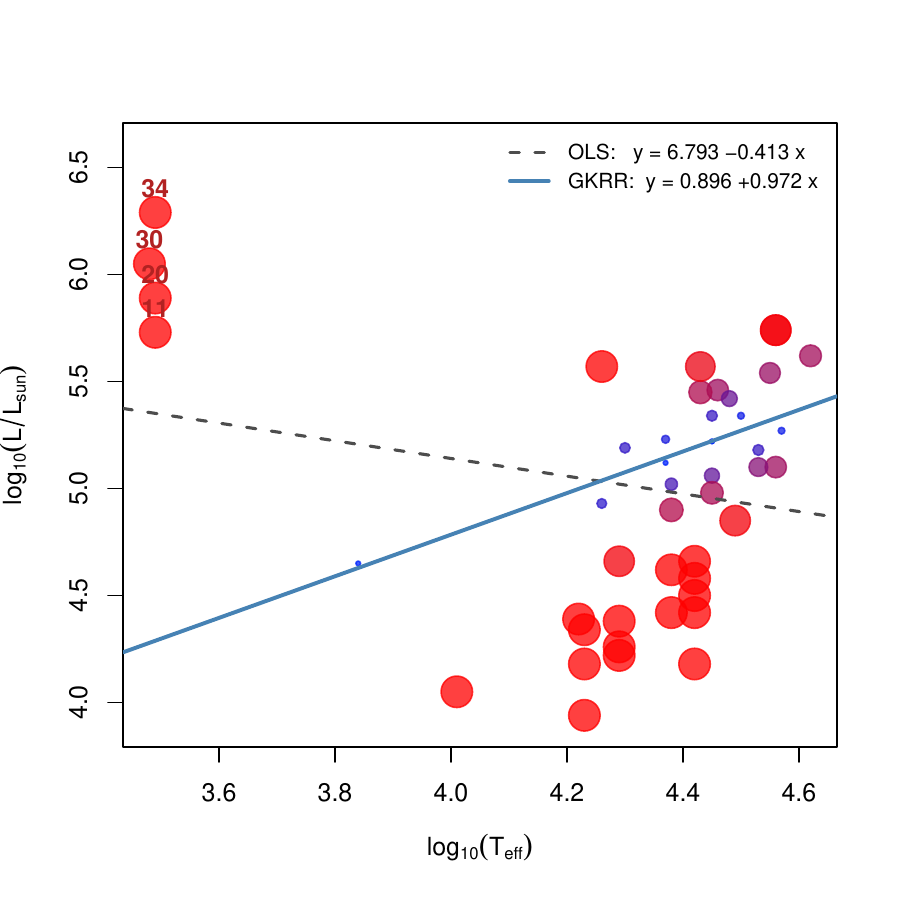}
\caption{\label{fig:stars_ols_x_gkrr}Comparison of the OLS fit (dashed grey line) versus the GKRReg fit (solid blue line) for the \code{stars\_cyg} data. Point size and colour are inversely proportional to the kernel weight $k_{ii}$ (blue/small = high weight, red/large = low weight); the four giant stars (observations~11, 20, 30, 34) are labelled.}
\end{figure}

The \code{stars\_cyg} dataset \citep{Humphreys:1978} is the Hertzsprung-Russell diagram of 47 stars in cluster CYG OB1. Four giant stars (observations~11, 20, 30, 34) lie well above the main sequence, acting as bad leverage points that severely distort OLS. Firstly, we present a brief comparison with the OLS estimates followed by the \code{gkrr} fit.
\begin{CodeChunk}
\begin{CodeInput}
R> data("stars_cyg")
R> fit_ols <- lm(log_light ~ log_temp, data = stars_cyg)
R> fit_stars <- gkrr(log_light ~ log_temp, data = stars_cyg,
+                    sigma_method = "s4")
R> round(rbind(OLS = coef(fit_ols),
+              GKRR = coef(fit_stars)), 4)
\end{CodeInput}
\begin{CodeOutput}
     (Intercept) log_temp
OLS       6.7935  -0.4133
GKRR      0.8960   0.9721
\end{CodeOutput}
\end{CodeChunk}
\begin{CodeChunk}
The effect of the four giant stars on the OLS slope estimate ($\hat\beta_1 = -0.41$) pulls the regression equation in the wrong direction. In the other hands, the GKRReg method downweights these leverage points recovers a \emph{positive} slope ($\hat\beta_1 = 0.97$), consistent with the expected main-sequence relationship in which hotter stars (higher \code{logtemp}) tend to be more luminous. Figure~\ref{fig:stars_ols_x_gkrr} plots the data with both regression lines overlaid, so the four giant stars stand out as large red points, while well-fitted main-sequence stars appear small and blue.
\begin{CodeInput}
R> summary(fit_stars)
\end{CodeInput}
\begin{CodeOutput}
Call:
gkrr(formula = log_light ~ log_temp, data = stars_cyg, sigma_method = "s4")
Residuals:
    Min      1Q  Median      3Q     Max 
-1.0678 -0.6103 -0.1194  0.1952  2.0015 
Coefficients:
            Estimate Std. Error CI 95% lower CI 95% upper p-value  
(Intercept)   0.8960     2.2460      -3.5061       5.2981  0.6900  
log_temp      0.9721     0.5562      -0.1181       2.0623  0.0805 .
---
Signif. codes:  0 '***' 0.001 '**' 0.01 '*' 0.05 '.' 0.1 ' ' 1
(Sandwich SE (asymptotic Wald z-test))
gamma^2: 0.05298  |  method: s4  |  iterations: 88  |  converged: TRUE
R-squared: -0.4700  |  Weighted R-squared: 0.5839
Kernel weights -- min: 0.0000  mean: 0.3167  max: 0.9999
Note: sandwich inference may be less reliable here (n = 47 (small sample);
43% of observations received near-zero kernel weight (< 0.01)).
  Consider bootstrap inference via boot = TRUE in gkrr() or
  summary(fit, boot = gkrr_boot(fit)).
\end{CodeOutput}
\end{CodeChunk}
Regarding inferential aspects of the GKRReg method, the sandwich-based test for this slope is only marginally significant ($p = 0.0805$), while the intercept is far from significant ($p = 0.690$). The wide confidence interval for the slope ($-0.12$ to $2.06$) reflects this uncertainty. As highlighted in the output note from the \code{gkrreg} package, 43\% of observations receive near-zero kernel weight, meaning that the effective sample size driving the fit is much smaller than the nominal $n = 47$. The package's own diagnostic note flags the asymptotic sandwich inference as potentially unreliable here and recommends bootstrap inference instead.

\begin{figure}[h!]
\centering
\includegraphics[width=0.6\textwidth]{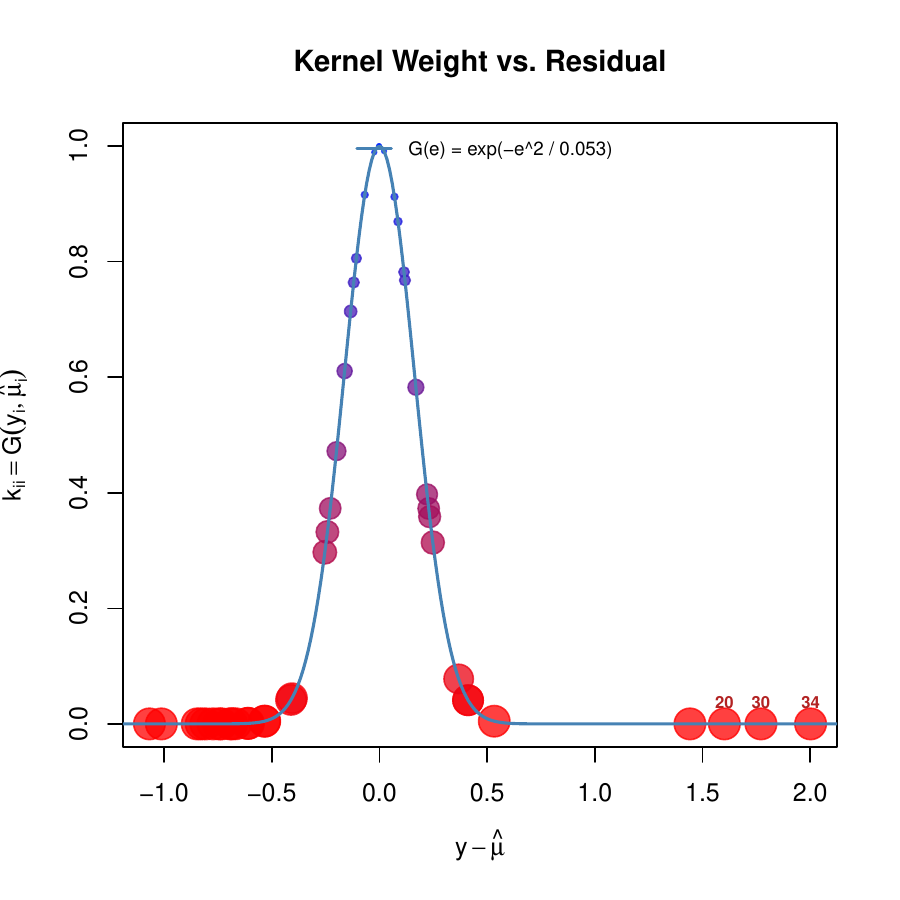}
\caption{\label{fig:stars_panel3}Panel~3 of \fct{plot.gkrr} for the
  \code{stars\_cyg} fit. The theoretical kernel curve
  $G(e)=\exp(-e^2/\hat\gamma^2)$ is overlaid.}
\end{figure}

The Figure~\ref{fig:stars_ols_x_gkrr} makes the mechanism behind the sign reversal explicit. The four giant stars sit far above the bulk of the data, and because they receive kernel weights near zero, the GKRReg line is essentially unaffected by them and instead follows the positive trend traced by the high-weight (blue) main-sequence stars. The OLS line, by contrast, is tilted toward the giants and ends up with the wrong sign.
\begin{CodeChunk}
\begin{CodeInput}
R> wt  <- fit_stars$weights
R> col <- rgb(1 - wt, 0.15 * wt, wt, alpha = 0.75)
R> cex <- 2.8 - 2.4 * wt
R> plot(stars_cyg$log_temp, stars_cyg$log_light,
+       pch = 19, col = col, cex = cex,
+       xlim = range(stars_cyg$log_temp), ylim = c(3.9, 6.6),
+       xlab = expression(log[10](T[eff])),
+       ylab = expression(log[10](L/L[sun])))
R> abline(fit_ols, lty = 2, lwd = 2, col = "grey30")
R> abline(fit_stars, lwd = 2.5, col = "steelblue")
R> text(stars_cyg$log_temp[c(11, 20, 30, 34)],
+       stars_cyg$log_light[c(11, 20, 30, 34)],
+       labels = c(11, 20, 30, 34), pos = 3, col = "firebrick", font = 2)
R> legend("topright", bty = "n", cex = 0.85,
+         legend = c(sprintf("OLS:   y = %.3f %+.3f x",
+                            coef(fit_ols)[1], coef(fit_ols)[2]),
+                    sprintf("GKRR:  y = %.3f %+.3f x",
+                            coef(fit_stars)[1], coef(fit_stars)[2])),
+         col = c("grey30", "steelblue"), lwd = c(2, 2.5), lty = c(2, 1))
\end{CodeInput}
\end{CodeChunk}
Finally, from another perspective, Figure~\ref{fig:stars_panel3} confirms that the same four giant stars (observations~11, 20, 30, 34) appear as large red points in the tail with near-zero weights, corroborating that GKRReg corrects the sign of the OLS slope.
\begin{CodeChunk}
\begin{CodeInput}
R> plot(fit_stars, which = 3, ask = FALSE)
\end{CodeInput}
\end{CodeChunk}


\section[Summary]{Summary} \label{sec:summary}

This paper establishes GKRReg as a redescending $M$-estimator and derives its
analytic sandwich variance estimator (HC0 class).
The key feature of this sandwich, relative to the classical OLS
case, is the correction term $(1-2 r_i^2/\hat\gamma^2)$ in the bread
matrix~\eqref{eq:Ahat}, which follows from differentiating the
Gaussian kernel with respect to $\boldsymbol\beta$.
A pairs bootstrap that re-estimates $\hat\gamma^2$ on each replicate
captures the variability of the algorithm and provides implicit
finite-sample corrections.
Three directions for future work are identified.
First, a GKRReg-HC3 estimator using the weighted leverage
$h_{\mathrm{GK},ii}$ from~\eqref{eq:hat_matrix} would improve small-sample
coverage without the computational cost of the bootstrap.
Second, formal theoretical results on the breakdown point of GKRReg as a
function of the chosen $\hat\gamma^2$ estimator would complement the
empirical findings of \citet{DeCarvalho+LimaNeto+Ferreira:2017}.
Third, the \code{"auto"} selection of $\hat\gamma^2$ by OOB bootstrap MSE
warrants a formal consistency analysis.
All methods are implemented in the \proglang{R} package \pkg{gkrreg},
freely available from CRAN.


\section*{Computational details}

The results in this paper were obtained using \proglang{R}~4.5.2 with the
\pkg{gkrreg}~0.4.0, \pkg{MASS}~7.3 \citep{Venables+Ripley:2002},
\pkg{sm}~2.2 \citep{Bowman+Azzalini:1997}, and
\pkg{robustbase}~0.99 \citep{Maechler+:2023} packages.
\proglang{R} and all packages are available from CRAN at
\url{https://CRAN.R-project.org/}.


\section*{Acknowledgments}

The authors thank Francisco de A.T.\ De Carvalho for the original
GKRReg method and for fruitful discussions. The second author thanks CNPq (National Council for Scientific and Technological Development) for their financial support.


\bibliographystyle{plainnat}
\bibliography{gkrreg_references}

\end{document}